\documentclass[11pt]{article} 
\usepackage{epsfig,a4}
\def\be{\begin{equation}}
\def\ee{\end{equation}}
\def\bea{\begin{eqnarray}}
\def\eea{\end{eqnarray}}

\def\C{{\rm\kern.24em
    \vrule width.02em height1.4ex depth-.05ex
    \kern-.26em C}}
\def\N{{\rm I\kern-.18em N}}
\def\R{{\rm I\kern-.21em R}}
\def\Z{{\rm\kern.26em
    \vrule width.02em height0.5ex depth 0ex
    \kern.04em
    \vrule width.02em height1.47ex depth-1ex
    \kern-.34em Z}}
\def\d{{\rm\kern.22em
    \vrule width.02em height1.0ex depth0ex
    \kern-.24em d}}

\newcommand\real{\mbox{Re}\,}
\newcommand\imag{\mbox{Im}\,}

\def\pbar{\bar{p}}

\def\kf{{\bf k}}

\def\C{{\rm\kern.24em
    \vrule width.02em height1.4ex depth-.05ex
    \kern-.26em C}}
\def\N{{\rm I\kern-.18em N}}
\def\Od{{\rm\kern.24em
    \vrule width.02em height1.45ex depth-.05ex
    \kern-.26em O}}
\def\oddi{{\rm\kern.24em
    \vrule width.02em height1.45ex depth-.05ex
    \kern-.26em O}}
\def\P{{\rm I\kern-.25em P}}
\def\R{{\rm I\kern-.21em R}}
\def\Z{{\rm\kern.26em
    \vrule width.02em height0.5ex depth 0ex
    \kern.04em
    \vrule width.02em height1.47ex depth-1ex
    \kern-.34em Z}}
\newcommand\spommi{{\mbox{\scriptsize\P}}}
\newcommand\soddi{{\mbox{\scriptsize\Od}}}
\newcommand\tpommi{{\mbox{\tiny\P}}}
\newcommand\toddi{{\mbox{\tiny\Od}}}
\newcommand\sbfkl{{\mbox{\scriptsize BFKL}}}

\newdimen\picraise
\newcommand\picbox[1]
{
  \setbox0=\hbox{\input{#1}}
  \picraise=-0.5\ht0 
  \advance\picraise by 0.5\dp0
  \advance\picraise by 3pt     
  \hbox{\raise\picraise \box0}
}
\newdimen\picraiset
\newcommand\picding[1]
{
  \setbox0=\hbox{\input{#1}}
  \picraiset=-0.5\ht0 
  \advance\picraiset by 0.5\dp0
  \advance\picraiset by -3pt  
  \hbox{\raise\picraiset \box0}
}
\newdimen\picraisehallo
\newcommand\pichallo[2]
{
  \setbox0=\hbox{\input{#1}}
  \picraisehallo=-0.5\ht0 
  \advance\picraisehallo by 0.5\dp0
  \advance\picraisehallo by -#2pt  
  \hbox{\raise\picraisehallo \box0}
}

\begin{document}
\begin{titlepage}
\begin{flushright}
HD-THEP-02-36\\
hep-ph/0403051
\\
\end{flushright}
\vfill
\begin{center}
\boldmath
{\LARGE{\bf The Perturbative Pomeron and the Odderon:}}
\\[.3cm]
{\LARGE{\bf Where can we find them?\,${}^*$}}
\unboldmath
\end{center}
\vspace{1.2cm}
\begin{center}
{\bf \Large
Carlo Ewerz
}
\end{center}
\vspace{.2cm}
\begin{center}
{\sl
Institut f\"ur Theoretische Physik, Universit\"at Heidelberg\\
Philosophenweg 16, D-69120 Heidelberg, Germany\\
\hspace{1cm}\\
email: carlo@thphys.uni-heidelberg.de}
\end{center}
\vfill
\begin{abstract}
\noindent
QCD predicts the existence of the perturbative 
Pomeron and of the Odderon. But both of them appear to be rather 
difficult to observe experimentally. We describe the experimental 
status of these two objects, discuss possible reasons for their elusive 
behavior, and point out promising search strategies.
\vfill
\end{abstract}
\vspace{5em}
\hrule width 5.cm
\vspace*{.5em}
{\small \noindent 
${}^*$ Invited talk at 26th Johns Hopkins Workshop on High-Energy Reactions, 
{\sl From the Standard Model to String Theory, from Colliders to 
Cosmic Rays}, Heidelberg, 1-3 August 2002. 
}
\end{titlepage}

\section{Introduction}

One of the most interesting problems in QCD is to understand 
the high energy limit of hadronic scattering processes. 
Already before the advent of QCD this problem had been 
widely studied in the framework of Regge theory, 
for a recent review see  \cite{Doschbook}. 
Based on the principles of analyticity, unitarity and 
Lorentz invariance of the scattering amplitude Regge 
theory gives an extremely successful phenomenological 
description of strong interactions in the Regge limit, 
i.\,e.\ in the limit of large center-of-mass energy 
$\sqrt{s}$ and relatively small momentum transfer 
$\sqrt{-t}$, the latter being chosen to be of the order of 
a hadronic mass scale. It is convenient in Regge theory to change 
from the squared energy $s$ to its conjugate variable, 
the complex angular momentum $\omega$, and via an integral 
transformation the scattering amplitude is  obtained as a 
function of $\omega$. Regge theory then relates the 
high energy behavior of hadronic scattering processes 
to the singularities of the scattering amplitude 
in the complex angular momentum plane, the Regge poles 
and Regge cuts. The leading contribution in the high energy 
limit is given by the rightmost singularity in the $\omega$-plane, 
the Pomeron. It can be interpreted as a $t$-channel exchange 
carrying vacuum quantum numbers between the scattering particles. 
In the framework of Regge theory the positions of the Regge singularities 
(together with their couplings to the scattering particles) 
are universal parameters which have to be determined from experimental 
data. Once known they can be used to predict cross sections for other 
processes. Regge theory is very successful in describing the wealth of 
available data on hadronic cross sections, including total cross sections, 
structure functions at small values of Bjorken-$x$ as well as diffractive 
processes. The high energy behavior of total hadronic cross sections 
for example is very well described by a Pomeron with intercept $1.09$ 
\cite{Groom:in}, leading to a slowly rising cross section 
$\sigma \sim s^{0.09}$. 
Obviously, one would like to derive Regge theory from QCD. But 
this is a difficult problem, mainly because scattering processes in 
the Regge limit are in general dominated by small momentum scales. 
A derivation of the Regge singularities from first principles would 
hence require a good understanding of nonperturbative QCD 
which we clearly do not have at present. 

In this talk I will address two issues which can help in making 
progress towards an understanding of Regge theory in terms of QCD. 
The first is the perturbative approach to the Pomeron which is a first 
step in the direction of deriving Regge singularities from QCD. The 
second is the Odderon, the partner of the Pomeron carrying negative 
charge parity quantum number. Although its existence is expected 
on the basis of our picture of high energy scattering in QCD, 
the Odderon has so far escaped unambiguous experimental detection. 
I will discuss mainly phenomenological aspects of the perturbative 
Pomeron and of the Odderon, in particular the presently available 
evidence for their existence. 

\section{The perturbative Pomeron}

The perturbative approach to high energy scattering is based on the 
resummation of large logarithms of the center--of--mass energy 
$\sqrt{s}$. The applicability of perturbation theory clearly requires 
that the value of the strong coupling constant can be assumed to be 
small in the scattering process under consideration. This means that 
the scattering process involves at least one large momentum scale. 
(Note that in a quantum field theory like QCD the running coupling 
constant always depends on the momentum of the particles, but not 
on the center--of--mass energy.) As is the case for any approximation 
scheme in physics one has to determine its range of applicability. 
As we will discuss below it turns out that in the case of the Pomeron 
there are additional effects which limit the applicability of perturbation 
theory. As a consequence the perturbative approach to the Pomeron 
can be used only in a rather limited number of scattering processes. 
Nevertheless, it is of enormous theoretical importance because it 
is, at least for the time being, the only rigorous way to understand 
the origin of Regge behavior of some scattering processes in terms of QCD. 

In the leading logarithmic approximation (LLA) one collects 
all diagrams of the perturbative series in which factors of the 
strong coupling constant $\alpha_s$ are accompanied by 
a logarithm of the energy $\sqrt{s}$. At high energies these 
logarithms can compensate the smallness of the strong coupling, 
and hence the LLA is characterized by 
\be
\alpha_s \ll 1 \,, \mbox{\hspace*{1.5cm}} 
\alpha_s \log s \sim 1
\,.
\ee
The resummation of all perturbative terms of the form 
$(\alpha_s \log s )^n$ in a two--particle scattering process 
was performed in \cite{Kuraev:fs,Balitsky:ic}, 
and the exchange in the $t$-channel of the resulting amplitude 
is known as the BFKL (Balitsky--Fadin--Kuraev--Lipatov) 
Pomeron. The BFKL equation describing that object is an 
evolution equation in energy. Equivalently, it can be interpreted 
as an evolution in rapidity along the $t$-channel or in longitudinal 
momentum of the real partons produced in the scattering process. 
The BFKL equation does however not include the effects 
of an evolution in the virtuality of the particles along the 
$t$-channel exchange. It is therefore strictly speaking only 
applicable to scattering processes which are completely 
dominated by only one hard momentum scale. The best 
processes for the study of the perturbative Pomeron are therefore 
scattering processes of two small color dipoles, and the size of the 
dipoles determines the hard scale necessary for the perturbative 
treatment. Theoretically, the best possible process would be 
the scattering of two heavy onia, which is however not accessible 
experimentally. Realistic processes which are close to this 
ideal situation are\footnote{Here I give only early references 
for the different processes. For further relevant references see papers 
referring to these.} Mueller--Navelet jets (i.\,e.\ forward jets) in 
proton--(anti)proton scattering \cite{Mueller:ey}, 
the production of hard forward jets in deep inelastic lepton--nucleon 
scattering\cite{Bartels:1996gr}, 
as well as the scattering of two virtual photons of the same 
or at least similar virtuality. In the latter case one can consider 
either quasi--diffractive processes like for example 
$\gamma^* \gamma^* \to J/\psi J/\psi$ \cite{Kwiecinski:1998sa}, 
or the total hadronic cross section in $\gamma^* \gamma^*$ collisions 
\cite{Bartels:1996ke,Brodsky:1996sg,Bartels:1997er,Brodsky:1997sd}. 

\begin{figure}
\begin{center}
\input{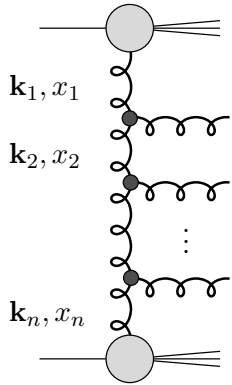}
\end{center}
\caption{Real gluon emissions from the $t$-channel gluon 
\label{emissions}}
\end{figure}

Before considering a concrete scattering process let us first 
discuss some basic properties of the BFKL Pomeron. 
The diagrams resummed in the LLA are of the form shown 
in figure \ref{emissions}. 
The figure shows a typical diagram contributing to the amplitude. 
A gluon is exchanged in the $t$-channel from which a number 
of real gluons can be emitted in the $s$-channel. In the LLA also 
virtual corrections are included which lead to the so--called 
reggeization of the $t$-channel gluon. In the process of reggeization 
the $t$-channel gluon becomes a more complicated object (it becomes 
a collective excitation of the gluon field rather than an elementary gluon), 
but for the present talk it will be sufficient to think of a gluon 
exchange in the $t$-channel. 
When the amplitude is squared in order to obtain the cross section 
one finds diagrams which exhibit the characteristic ladder structure 
of the perturbative Pomeron. The gluons along the ladder are 
strongly ordered in rapidity, or alternatively in longitudinal momentum 
fraction $x_i$. Only these strongly ordered configurations give rise 
to a logarithm of the energy $\sqrt{s}$ for each factor of the coupling 
constant $\alpha_s$. The transverse momenta $\kf_i^2$ 
of the emitted gluons, on the other hand, are not ordered. 
This has to be contrasted with the situation 
in DGLAP evolution \cite{Gribov:ri,Altarelli:1977zs,Dokshitzer:sg} 
where similar diagrams occur, but with the transverse momenta 
strongly ordered along the ladder. 
The BFKL equation is usually written as an integral equation in 
which the integral kernel represents a rung of the ladder, i.\,e.\ 
the emission of a real gluon. Solving the BFKL equation one finds 
for the energy dependence of the cross section in the LLA 
a powerlike growth, $\sigma \sim s^{\omega_{\mbox{\tiny BFKL}}}$, 
with the BFKL exponent given by 
\be
\label{BFKLexponent}
\omega_{\mbox{\scriptsize BFKL}} = \frac{\alpha_s N_c}{\pi} \, 4 \log 2 
\,.
\ee
It is of the order of $0.5$ when a typical value of $0.2$ is chosen 
for the strong coupling constant $\alpha_s$. 
The so--called Pomeron 
intercept is hence given by 
$\alpha_\spommi = 1 + \omega_{\mbox{\scriptsize BFKL}}$. 
The LLA does not include the running of the coupling constant $\alpha_s$. 
Strictly speaking one hence has to use a fixed $\alpha_s$ in this approximation, 
although it is widely believed that replacing a fixed value by the running 
coupling is an improvement. 
Note that the energy dependence of the cross section in the LLA depends 
exponentially (read: very strongly) on the coupling constant $\alpha_s$. Given the 
above formula this is a rather trivial observation. Nevertheless, it constitutes 
a rather important limitation for making accurate predictions within 
this approximation scheme because the determination of the appropriate 
value of $\alpha_s$, or equivalently of the relevant momentum scale, 
is often rather difficult in practice. Due to that any prediction made 
in the framework of the LLA has an unavoidable uncertainty which, 
although its origin is trivial, can be quite large. The only way to avoid 
this problem at least partially is to go to the next--to--leading logarithmic 
approximation (NLLA) in which running coupling effects naturally 
occur. We will discuss that approximation further below. 

An advantage of the BFKL equation is that it can be solved analytically. 
The solution represents ladder diagrams with arbitrarily many rungs. 
An important point to note here is that in the LLA the gluons emitted 
from the $t$-channel gluon can be produced without any cost in energy. 
Hence energy conservation is violated at the emission vertices. This is not 
because the authors had not been informed about energy conservation 
in Nature, but is simply an outcome of the approximation scheme. 
In the sense of leading logarithms of the energy the effect of the 
correct kinematics of the vertices is a subleading effect. The formally 
subleading effect of energy--momentum conservation 
at the emission vertices has been studied by implementing a so--called 
consistency constraint in the BFKL equation \cite{Kwiecinski:1996td}. 
In a Monte Carlo study of the correspondingly modified BFKL evolution 
it was found that the constraint considerably reduces the growth 
of the cross section with the energy \cite{Orr:1998rw}. 
Physically this means that the production of a real gluon requires 
some amount of energy, and realistically we should not expect 
arbitrarily large numbers of gluons to be produced. We will come back 
to this effect further below. 

\begin{figure}
\begin{center}
\input{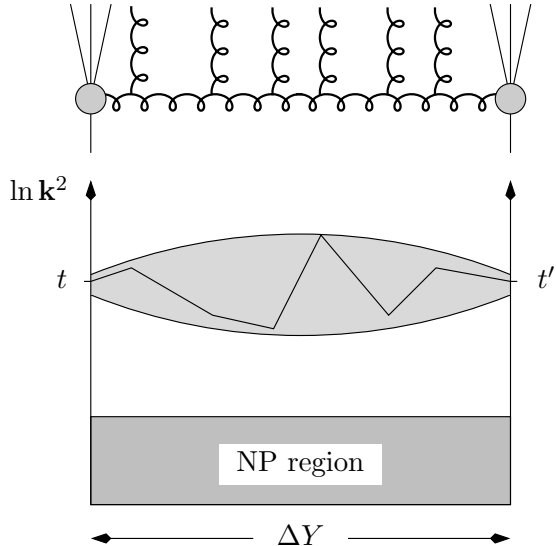}
\end{center}
\caption{Diffusion of transverse momenta in the BFKL Pomeron
\label{figdiffusion}}
\end{figure}

Another phenomenologically important property of the BFKL Pomeron 
follows immediately from the fact that the transverse momenta $\kf_i^2$ 
of the gluons are not ordered along the ladder. As a consequence there 
is nothing that prevents the transverse momenta from becoming 
arbitrarily small. One in fact finds that the gluon emissions along the 
ladder lead to a random walk in $\log \kf_i^2$. 
The resulting probability distribution of momenta along the ladder 
resembles a diffusion process. This is illustrated in figure 
\ref{figdiffusion}. 
Here the horizontal axis shows the rapidity interval 
between the ends of the ladder, $\Delta Y \sim \log(s)$. 
The vertical axis shows the logarithm of the transverse momentum 
which is fixed at the ends of the ladder at values $t=t'$ determined 
by the typical momentum scales of the external particles. 
The resulting probability distribution is known as the Bartels cigar. 
The exact shape of the cigar depends on the external momentum 
scales and on the rapidity interval available for evolution 
\cite{Bartels:1993du,Bartels:1995yk}. 
With increasing energy in the scattering process the momentum 
distribution becomes wider in the middle. Therefore some contribution 
from the nonperturbative region of small momenta cannot be 
avoided at very high energies even when the external momenta are 
chosen very large. But in situations in which this contribution is large 
the whole perturbative description using the BFKL Pomeron 
is no longer applicable. The situation is particularly severe if the 
external particles provide small momentum scales, and this 
already indicates that the BFKL Pomeron cannot be used for 
the description of structure functions of the nucleon since there 
one end of the ladder resides completely in the nonperturbative 
region. 

\begin{figure}
\vspace*{0.4cm}
\centering
\includegraphics[width=0.4\textwidth,clip]{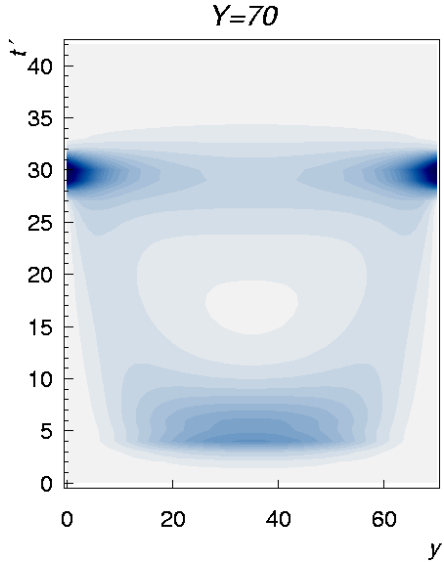}
\caption{Transition from cigar--type evolution to
tunneling--type evolution, figure from \protect\cite{Ciafaloni:2002xk}
\label{figalien}}
\end{figure}

Interestingly, the situation becomes even worse when the coupling 
constant $\alpha_s$ is assumed to run as a function of the gluon 
momenta along the ladder \cite{Ciafaloni:2002xk}. In that case 
the probability distribution of the transverse gluon momenta is no 
longer symmetric in the vertical direction. 
Instead emissions with 
smaller momenta become more likely as the coupling constant $\alpha_s$ 
is larger at smaller momenta. Due to that the distribution takes a 
banana shape rather than the cigar shape. At very large energies, 
or equivalently large rapidity intervals for evolution, even a tunneling 
transition takes place. Then the first emission brings the gluon into 
the infrared region where it stays until the last step of the evolution. 
A numerical simulation of such a situation is shown in figure 
\ref{figalien}. 
In such a situation the perturbative description breaks 
down completely and the process is determined completely by the 
soft Pomeron. Note that even for large external momentum scales 
$t$ this eventually happens as the energy becomes very large. 
The problem of diffusion of the transverse momenta is an important 
limitation for the applicability of the BFKL Pomeron. Fortunately, 
one can find out via numerical simulation whether a given process 
involves a large contribution from the nonperturbative region or not. 
In addition, one can choose suitable cuts in a number of scattering 
processes such that the external momenta are large enough to 
suppress the diffusion into the infrared. We will now see an 
example for such a process. 

\begin{figure}
\begin{center}
\input{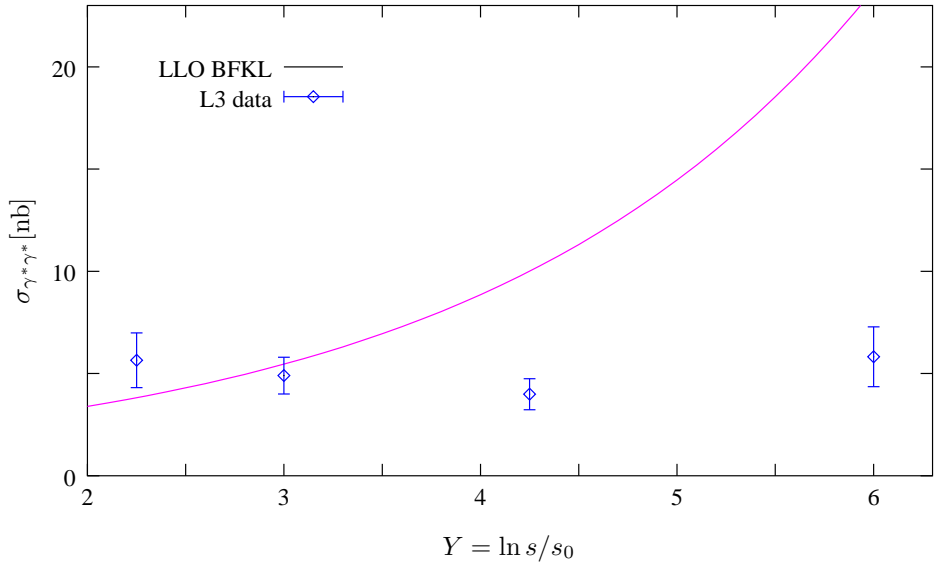}
\end{center}
\caption{Data for the total hadronic cross section in $\gamma^* \gamma^*$ 
collisions at LEP2 as measured by the L3 collaboration 
\protect\cite{Achard:2001kr} compared to the LLO BFKL prediction 
of \protect\cite{Bartels:2000sk}
\label{gagal3fig}}
\end{figure}

Let us now consider the total hadronic cross section of virtual 
photon--photon scattering as a specific example of a process in 
which one would expect to see effects of the perturbative Pomeron, 
namely a rise of the cross section with energy. As already 
mentioned in the introduction this process is one of the best 
possible probes of BFKL dynamics at least from a theoretical 
point of view. This process has been studied in $e^+e^-$ 
collisions at LEP. Here one selects so--called double--tagged 
events in which both the scattered electron and positron are 
detected and the virtualities of the photons emitted from the 
two can be reconstructed. 
If the two photon virtualities are chosen large 
enough and of the same order of magnitude the process is in 
fact determined by only one hard momentum scale. 
This process was first studied in 
\cite{Bartels:1996ke,Brodsky:1996sg,Bartels:1997er,Brodsky:1997sd}, 
later on these studies were refined in various ways, including 
attempts at including NLLO effects. Here we show results obtained 
in \cite{Bartels:2000sk} using the LLA. That calculation in 
particular includes the effect of the charm quark mass which is 
quite important for this process. 
The corresponding theoretical expectations based on the LLA are shown in 
figure \ref{gagal3fig}. 
The cross section has been measured by the 
OPAL and L3 collaborations at LEP \cite{Abbiendi:2001tv,Achard:2001kr}, 
and figure \ref{gagal3fig} shows the L3 data from \cite{Achard:2001kr}. 
There are still quite a few uncertainties in the theoretical expectation, 
in particular corresponding to the appropriate choice of $\alpha_s$ and 
of the fixed energy scale $s_0$. This scale $s_0$ should be a characteristic 
scale for the process, here it is chosen as the geometric mean of the 
two photon virtualities. (Strictly speaking it can only be fixed in a NLLA 
calculation.) But these uncertainties do by no means affect the obvious 
conclusion that the data remain far below the BFKL expectation 
based on the LLA. At least at energies accessible at LEP a strong rise 
of the cross section is clearly not visible. 

\begin{figure}
\vspace*{0.4cm}
\centering
\includegraphics[width=0.6\textwidth,clip]{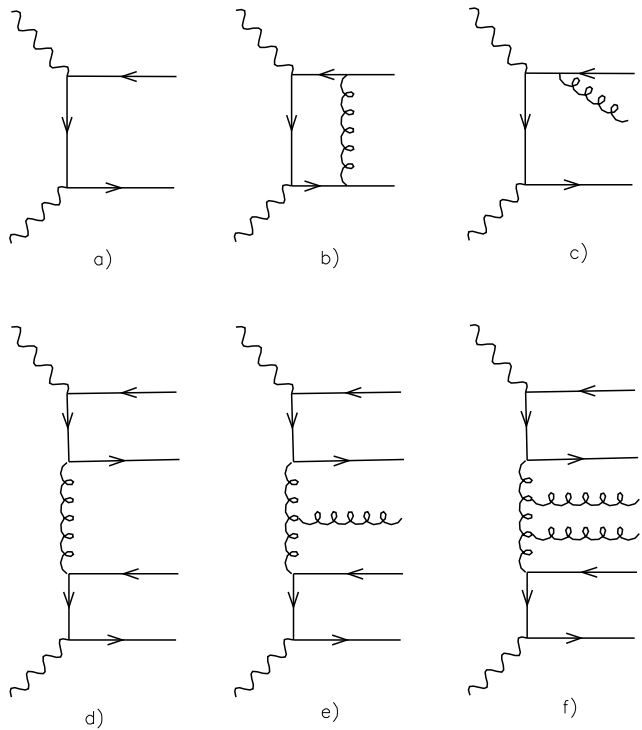}
\caption{Typical diagrams for the production of hadrons in 
$\gamma^* \gamma^*$ collisions, figure from 
\protect\cite{DelDuca:2002qt}\label{diagmaltoni}}
\end{figure}

\begin{figure}[ht]
\vspace*{0.4cm}
\centering
\includegraphics[width=0.75\textwidth,clip]{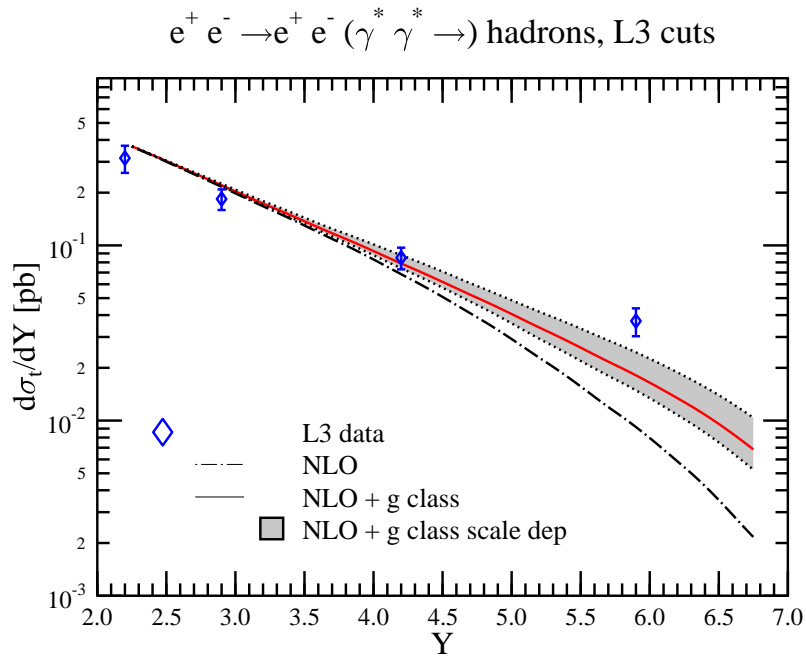}
\caption{Prediction for the total hadronic cross section for 
the process $e^+e^- \to e^+e^- X$ in NLO (dot-dashed line), 
and NLO plus diagrams of type d) in figure \protect\ref{diagmaltoni} 
(solid line), 
together with the data from \protect\cite{Achard:2001kr}, 
figure from \protect\cite{DelDuca:2002qt}\label{figmaltoni}}
\end{figure}

This result raises the question whether the data can be explained completely 
without any resummation of large logarithms of the energy. In order to 
answer this question the data have been compared with a fixed order 
calculation in \cite{DelDuca:2002qt}. The calculation performed there 
is done in next--to--leading order (NLO) and hence includes diagrams of the 
type a)--c) in figure \ref{diagmaltoni}. 
At high energies one expects that diagrams of the type d)--f) 
become increasingly important since they contain the 
enhancement due to logarithms of $\sqrt{s}$. The diagrams of 
type a)--c) on the other hand involve an exchange of a fermion 
in the $t$-channel. It is known that these diagrams are necessarily 
suppressed at high energies by a power of $\sqrt{s}$. 
At large energies one therefore expects at least the diagram of 
type d) to become important, which does not have any energy 
dependence. The actual BFKL type enhancement only starts with 
the diagram of type e). The result of the NLO calculation is shown 
in figure \ref{figmaltoni}, here on the level of the $e^+e^-$ cross 
section. 
The pure NLO curve is well below the data especially at large 
energies. If one includes the diagrams of type d) one obtains the 
solid curve which is only slightly below the data. The calculation 
in \cite{DelDuca:2002qt} is performed with four massless quark 
flavors. In reality the mass of the charm quark suppresses the 
production rate of charm quarks in this process, 
and according to \cite{Bartels:2000sk} the calculation with a 
massless charm quark overestimates the actual cross section 
by about 15\,\%. Taking this effect into account one concludes 
from figure \ref{figmaltoni} that the data point at the highest 
energy actually exceeds the expectation of a fixed order calculation 
for this process. But the enhancement at high energy is by far 
not as large as predicted by BFKL resummation in LLA. 
As we will explain further below a consistent calculation of this 
prediction in next--to--leading logarithmic accuracy is not available 
at present. 

Also in other processes in which the perturbative Pomeron 
can be looked for the situation is similar. An unambiguous sign 
of the BFKL Pomeron has not yet been found. One often finds an 
enhancement at high energies, but it is usually much smaller 
than one would have expected in LLA. 
We are thus led to the conclusion that the LLA is not applicable 
in the situations that are experimentally accessible at present. 
In each case there are different problems, but one key problem 
is common to all these cases. This issue is closely related to 
the problem of the violation of energy conservation in the LLA 
that we have already discussed. The analytic solution of the 
BFKL equation contains diagrams with arbitrarily many 
gluon rungs. In reality, however, the emission of a gluon into 
the final state requires a certain amount of energy. Our experience 
with deep inelastic scattering at HERA tells us that this amount 
of energy can roughly be estimated to correspond to one unit 
in rapidity. Events corresponding to the highest available energies 
at HERA of LEP for example span a rapidity range of about 5 or 6 
units. In addition we have to take into account that also the breakup 
of the incoming particles, for example the transformation of a virtual 
photon into a quark--antiquark pair, takes up at least one unit in rapidity. 
In total we should therefore expect that in reality only two or three 
gluons can be emitted. It is therefore not at all surprising that at 
these energies a fixed order calculation comes close to the 
measured cross section. 

\begin{figure}
\vspace*{0.4cm}
\centering
\includegraphics[width=0.6\textwidth,clip]{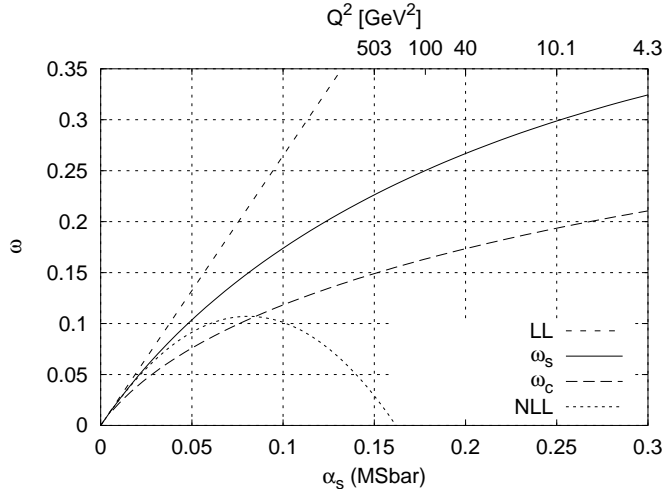}
\caption{BFKL exponent as a function of the coupling 
constant $\alpha_s$ using a renormalization group 
improvement of the NLLO BFKL equation, figure from 
\protect\cite{Ciafaloni:1999yw}\label{figomcrit}}
\end{figure}

As already mentioned above the problem of energy conservation 
at the emission vertices is closely related to next--to--leading 
logarithmic corrections to the BFKL equation. The derivation 
of the BFKL equation in NLLA is much more difficult than 
the LLA version. In an effort that lasted for almost ten years 
that problem of including terms of 
the order $\alpha_s (\alpha_s \log s)^n$ has been solved, 
see \cite{Fadin:1998py,Ciafaloni:1998gs} and references therein. 
The corrections were found to 
be rather large, giving for the characteristic 
exponent $\omega_\sbfkl$ of the energy dependence 
\be
\label{omnlla}
\omega_{\sbfkl} \simeq 2.65 \, \alpha_s (1 - 6.18 \,\alpha_s)
\,,
\ee
where the first term corresponds to the exponent in LLA. 
The large correction indicates a poor convergence 
of the perturbative series. Initially this 
result led to serious doubts about the BFKL approach 
in NLLA. These doubts have been considerably weakened 
after the problem was subsequently studied in more detail. 
The large corrections were found to originate from 
collinear divergences due to the emission of real gluons 
that are close to each other in rapidity. 
Several methods have been proposed to circumvent 
this problem, among them 
the application of a Brodsky--Lepage--Mackenzie 
(BLM) scale setting procedure \cite{Brodsky:1998kn}, 
a method to veto the emission of gluon pairs 
close in rapidity in a Monte Carlo implementation 
of the BFKL equation \cite{Schmidt:1999mz}, 
and a renormalization 
group improvement of the BFKL equation resumming 
additional large logarithms of the transverse 
momentum \cite{Ciafaloni:1999yw}. The result of the 
latter procedure is shown in figure \ref{figomcrit}. 
The figure shows the dependence of the BFKL exponent 
$\omega_{\sbfkl}$ on the 
strong coupling $\alpha_s$. The short-dashed line shows the 
exponent in LLA, the dotted line is the exponent according 
to NLLA without resummation as given in (\ref{omnlla}). 
The solid line is the result obtained after applying the renormalization 
group improvement. 
All of the methods mentioned above lead to stable results for the 
BFKL exponent, although the precise values differ slightly for 
the different methods. 
With those improvements the BFKL 
equation in NLLA is now widely considered a reasonable 
approximation scheme. For typical values of $\alpha_s$ 
around $0.2$ one now obtains a typical exponent for 
the energy dependence of about 
$0.2$ to $0.3$, to be compared with the LLA value of $0.5$. 

\begin{figure}
\begin{center}
\input{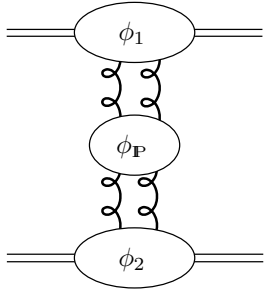}
\end{center}
\caption{Factorization of the perturbative Pomeron amplitude 
in the high energy limit \label{figpomfact}}
\end{figure}

An obvious question arising here is whether the BFKL Pomeron 
in NLLA can describe the data for the total hadronic cross section 
in virtual photon collisions discussed above. At present the answer 
to that question is unknown. At high energies the cross section 
for a given process can be factorized in the form shown in 
figure \ref{figpomfact}. 
Obviously, the cross section 
involves not only the Pomeron amplitude $\phi_\omega$, 
but also the impact factors $\phi_{1,2}$ which describe the coupling 
of the Pomeron to the external particles. The Pomeron amplitude 
is known in NLLA, but a consistent calculation requires also the 
impact factors in that approximation. At the time of this conference 
they are not yet fully known. So far, the real and virtual parts 
of these impact factors have been calculated separately, 
but there remain a number of phase space integrals 
to be done, see \cite{Bartels:2000gt,Fadin:2001ap} and references therein. 

We should emphasize that in spite of all the uncertainties and problems 
mentioned above it is a rather firm prediction of perturbative QCD that the 
cross section of two small color dipoles eventually rises with the energy. 
Practically, however, there are many caveats in applying the 
LLA or NLLA to a specific process at a given energy. It is therefore 
very difficult to isolate the perturbative Pomeron experimentally. 
Personally, I would guess that the total hadronic cross section 
in virtual photon collisions at a future Linear Collider will offer 
the best chances, in particular due to the larger energy and 
luminosity. 
Notwithstanding the problems of observing the perturbative 
Pomeron the resummation of logarithms of the energy is 
extremely valuable for studying perturbative QCD, in particular 
because it gives us information about the small-$x$ anomalous 
dimension of the gluon. 

\section{The Odderon}

The Odderon is the $C=-1$ partner of the Pomeron. It is 
defined as the leading contribution to the odd--under--crossing 
amplitude at high energies with an intercept $\alpha_\soddi$ 
close to one. Due to its negative charge parity the Odderon 
gives a contribution to the difference of 
particle--particle and  particle--antiparticle cross sections. 
The Odderon was introduced in 
the framework of Regge theory almost thirty years ago 
in \cite{Lukaszuk:1973nt}, but was for a long time considered a 
heretic and doubtful concept. Initially, one of the reasons for this was 
that the widely known Pomeranchuk theorem \cite{Pomeranchuk} 
states that the cross sections for particle--particle 
and particle--antiparticle scattering become equal at high energies. 
For the specific example of $pp$ and $p\bar{p}$ scattering this means 
\be
\label{pomtheorem}
\Delta \sigma = \sigma_T^{\bar{p}p} - \sigma_T^{pp} 
\mathop{\longrightarrow}_{s \to \infty}
0
\,.
\ee
However, the proof of this theorem {\sl assumes} that the odd--under 
crossing amplitude vanishes. Without that assumption, one can show that 
\be
\label{genpomtheorem}
\frac{\sigma_T^{\bar{p}p}}{\sigma_T^{pp} } 
\, \mathop{\longrightarrow}_{s \to \infty} \,
1
\,,
\ee
which does not contradict the existence of the Odderon as can be seen in 
the following toy example for a possible behavior of the two cross sections: 
\bea
\sigma_T^{pp} &=& A \log^2 s + B \log s + C \\
\sigma_T^{\bar{p}p} &=& A \log^2 s + B' \log s + C'
\,.
\eea
Clearly, if $B \neq B'$ the general Pomeranchuk theorem 
(\ref{genpomtheorem}) is satisfied, but the original 
Pomeranchuk theorem (\ref{pomtheorem}) 
is violated, and instead in this particular example one even has 
$|\Delta \sigma | \to \infty$ for 
$s \to \infty$. 

The Odderon received more attention after it had been observed that 
in QCD it can be build as a state of three gluons in a symmetric 
color state. A Pomeron in the simplest picture consists of a two--gluon 
exchange, and at least the nonperturbative version of the Pomeron clearly 
exists and has been observed in many scattering processes. 
There is a priori no reason why an exchange of three gluons should not 
exist. This simple observation strongly suggests that an Odderon of some 
kind, be it perturbative or nonperturbative, should occur in high energy 
scattering. In the following I will briefly mention some 
basic facts about the perturbative Odderon and then concentrate 
on phenomenological issues. For a more detailed  review on the 
Odderon see \cite{Ewerz:2003xi}. 

\begin{figure}[ht]
\begin{center}
\input{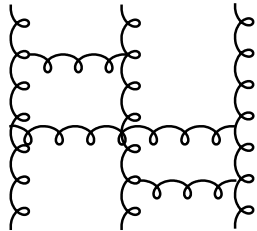}
\end{center}
\caption{Typical ladder diagram contributing to the perturbative 
Odderon\label{figoddi}}
\end{figure}

In perturbative QCD the Odderon is described by the 
Bartels--Kwieci{\'n}ski--Prasza{\l}o\-wicz (BKP) equation 
\cite{Bartels:1980pe,Kwiecinski:1980wb}. In analogy to the BFKL 
equation it resums the leading logarithms of the energy $\sqrt{s}$ 
for a state of three (reggeized) gluons in the $t$-channel. The 
corresponding diagrams again have a ladder structure with pairwise 
interactions of the three gluons as is illustrated in figure \ref{figoddi}. 
Remarkably, the BKP equation has a hidden conserved charge 
\cite{Lipatov:1993qn,LipatovPadova} and is therefore a completely 
integrable system. It in fact turns out that it is equivalent to the 
XXX Heisenberg model consisting of three sites with noncompact 
$\mbox{SL}(2,\C)$ spin $s=0$ \cite{Lipatov:1993yb,Faddeev:1994zg}. 

In the past years explicit solutions of the BKP equation have been 
found, and it is likely that by now all solutions of the BKP equation 
are known. The first type of solution was found in \cite{Janik:1998xj} 
and is called the Janik--Wosiek (JW) solution. 
Its intercept is was found to be
\be
\alpha_\soddi = 1 - 0.24717 \,\frac{\alpha_s N_c}{\pi}\,,
\ee
and for a typical value $\alpha_s \simeq 0.2$ this yields 
$\alpha_\soddi \simeq 0.96$. 
Consequently one obtains an almost flat energy dependence of 
$s^{\alpha_\toddi -1}$ for the Odderon exchange in the case 
of the JW solution. Another type of solution of the BKP equation, the 
Bartels--Lipatov--Vacca (BLV) solution, was found in 
\cite{Bartels:1999yt}. This solution has been constructed explicitly in 
terms of eigenfunctions of the BFKL equation. The intercept of 
the BLV Odderon solution is exactly $\alpha_\soddi =1$. The 
exchange of the BLV Odderon should hence persist to very high 
energies. 
Naively one would conclude here that the leading contribution 
to the Odderon comes from the solution with the highest intercept, 
i.\,e.\ the BLV solution. However, the situation is more complicated.  
The complication comes due to the fact that the two types of solutions 
exhibit a very different behavior concerning the coupling to 
external particles. The JW solution for example does not couple 
(at least in leading logarithmic order) to the phenomenologically 
interesting photon-$\eta_c$ impact factor. The BLV solution 
on the other hand does couple to this impact factor. As a consequence 
only the BLV solution contributes to the quasidiffractive process 
$\gamma^* \gamma^* \to \eta_c \eta_c$, for instance. In other processes, 
for example in proton--proton scattering at large $t$ both solutions are 
expected to contribute. In such processes the relative importance 
of the two solutions is mainly determined by the coupling to the 
external particles because their intercepts are almost equal. 
There are strong indications that in most processes of phenomenological 
interest the BLV solution gives the leading contribution, and sometimes 
the only one. 

We have seen that at least in perturbative QCD the occurrence of 
the Odderon is very natural. There is no obvious reason why processes 
involving Odderon exchange should have a very small cross section. 
This is supported by the fact that the perturbative Odderon intercept 
is exactly one (or only slightly below one for the JW solution). 
The situation is less clear when one turns to processes in which 
perturbation theory is not applicable. Only very little is known about 
the soft or nonperturbative Odderon. But also 
at low momentum scales there is no obvious reason for the absence 
of the Odderon. In the contrary, due to the larger value of the 
coupling constant at small momenta a three--gluon (Odderon) 
exchange should be even less suppressed with respect to the 
two--gluon (Pomeron) exchange. Our picture of high energy 
scattering based on gluon exchange, be it perturbative or 
nonperturbative, hence strongly suggests that Odderon exchange 
should exist and should lead to sizable cross sections. 
Reality seems to be different. 

\begin{figure}
\begin{center}
\input{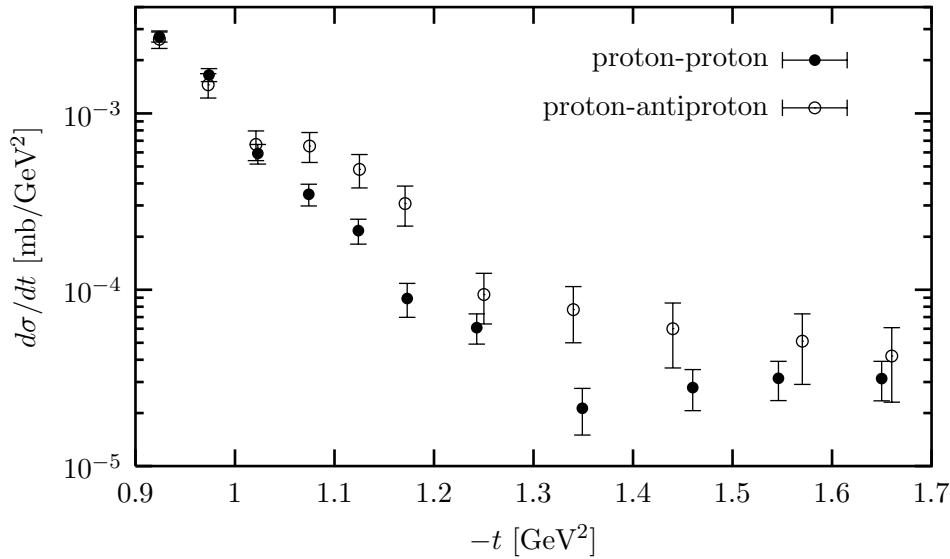}
\end{center}
\caption{Differential cross section for elastic $pp$ and $p\bar{p}$
scattering in the dip region for $\sqrt{s}=53\,\mbox{GeV}$;
data from \protect\cite{Breakstone:1985pe}, figure 
from \cite{Ewerz:2003xi}
\label{fig:dipdiff}}
\end{figure}

So far the only evidence for the Odderon has been found in the 
difference of the differential cross sections for elastic $pp$ and 
$p\bar{p}$ scattering 
\be
\frac{d\sigma^{\bar{p}p}_{el}}{dt} - \frac{d\sigma^{pp}_{el}}{dt}
\ee 
in the dip region around $t \simeq -1.3\,\mbox{GeV}^2$. 
Figure \ref{fig:dipdiff} shows the corresponding data taken 
at $\sqrt{s}=53\,\mbox{GeV}$ at the CERN ISR 
\cite{Breakstone:1985pe}. 
The $pp$ data show a dip whereas the $p\bar{p}$ data only 
flatten off at the same momentum transfer. Such a difference is 
a typical sign of an exchange carrying negative charge parity. 
It can be shown that pure reggeon exchange is not sufficient 
to produce the observed difference, and hence the data indicate 
an Odderon contribution. But there are two caveats here. 
First, the statistics of the data shown in figure \ref{fig:dipdiff} 
is rather low. The $p\bar{p}$ data have in fact been taken only during 
the last week of running of the ISR. 
Unfortunately, there is no other energy at which we have 
data for both $pp$ and $p\bar{p}$ elastic scattering. 
Comparing the two at largely different energies necessarily 
requires theoretical models for the description of the data 
which are mostly in the soft (nonperturbative) region. 
This brings us to the second caveat. The available models 
for the elastic scattering data involve 
a large number of exchanges in addition to the Odderon, 
like Pomeron, double Pomeron, reggeon etc. Accordingly, 
these models have typically around 20 to 40 parameters. With the 
available data these simply cannot be determined precisely 
enough to cleanly identify the Odderon contribution. 
We hope that elastic scattering data from RHIC will 
improve this situation in the future. 
It is worth noting though that almost all fits require an 
Odderon contribution of some kind in order to describe 
the available data. 

Although the models necessary for the description of the 
differential cross sections of elastic $pp$ and $p\bar{p}$ 
scattering do not allow one to extract the Odderon 
contribution with sufficient precision it is still possible 
to gain interesting information on the Odderon from 
these data. In \cite{Dosch:2002ai} for example the 
coupling of the Odderon to the proton was investigated 
and found to depend strongly on the internal structure 
of the proton. As a framework the Regge description of 
the elastic scattering data due to Donnachie and Landshoff (DL) 
\cite{Donnachie:hf} is used. Then the Odderon contribution 
in that fit is replaced by a perturbative three--gluon exchange, 
hoping that in the dip region the momentum transfer is 
still large enough to allow for a perturbative picture, 
at least in some reasonable approximation. 
One can then use different models for the coupling of the 
Odderon to the proton and try to determine their 
parameters. It turns out that in this way one can, within 
a given model, constrain the parameters very strongly. 
Therefore it is likely that the qualitative results are rather 
independent of the parametrization chosen as a framework. 

\begin{figure}
\vspace*{0.4cm}
\centering
\includegraphics[width=0.5\textwidth,clip]{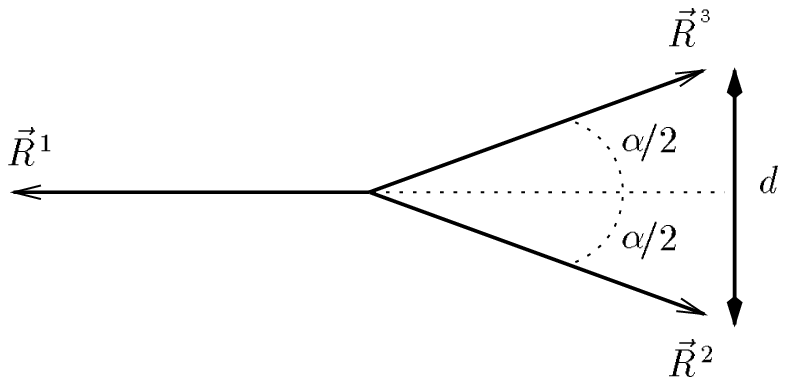}
\caption{Definition of the angle $\alpha$ characterizing the 
proton configuration, figure from \protect\cite{Dosch:2002ai}
\label{star}}
\end{figure}

In \cite{Dosch:2002ai} this has been done for 
two impact factors that had been proposed in \cite{Fukugita:1978fe} 
and \cite{Levin:gg}, respectively, and also for a simple 
geometric model for the transverse structure of the proton. 
The two impact factors contain as a main parameter the value of 
the coupling $\alpha_s$. 
The geometric model assumes a distribution of the three 
valence quarks in the proton of the form shown in figure \ref{star}. 
Here the main parameter is the size $d$ of a pair of quarks, 
whereas the radius of the configuration is essentially fixed 
by the electromagnetic radius of the proton. In this model 
a possible quark--diquark structure of the proton would 
simply correspond to a small value of $d$ (or a small angle 
$\alpha$). 
\begin{figure}
\begin{center}
\input{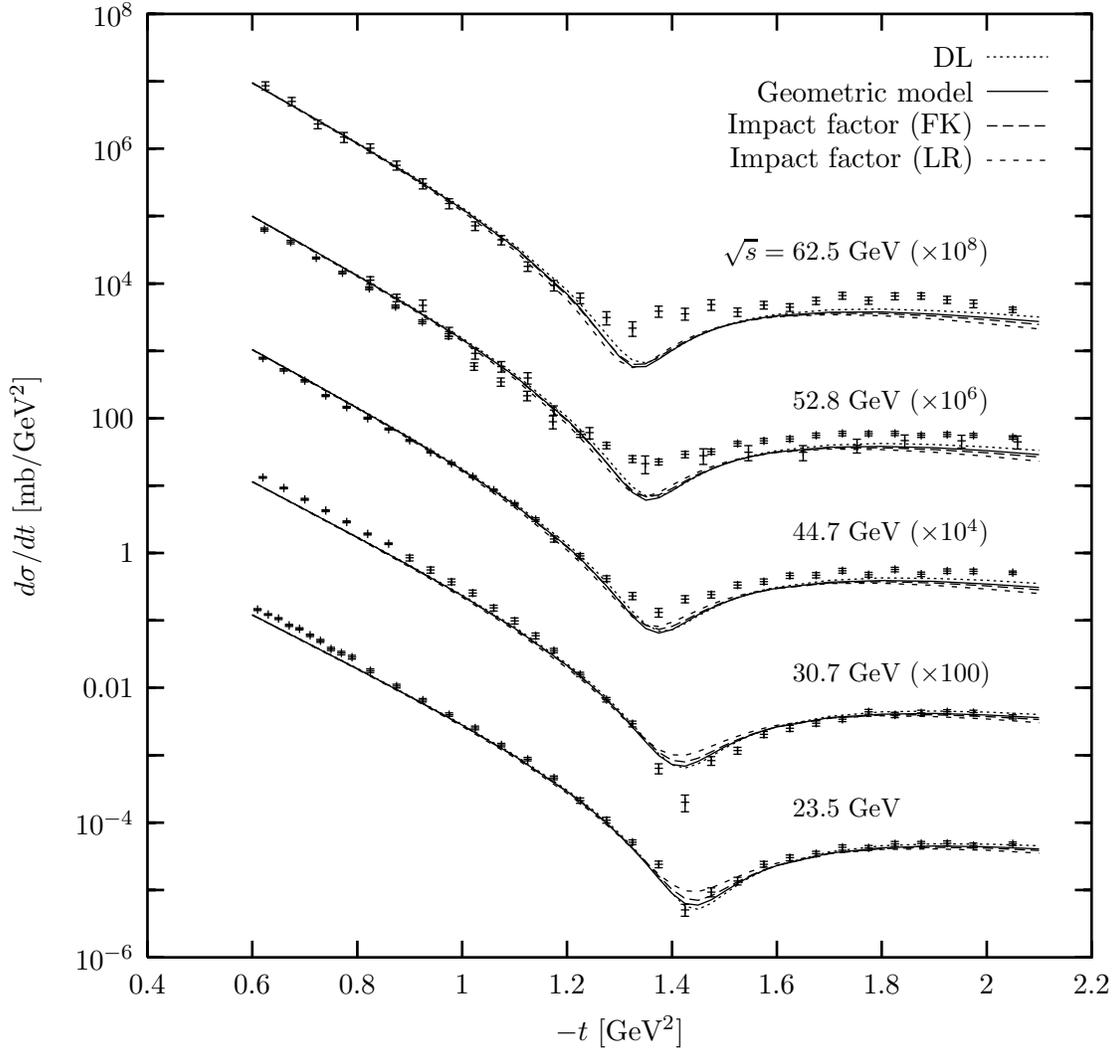}
\end{center}
\caption{Differential cross section for elastic $pp$ scattering
calculated using different couplings of the Odderon to the proton:
the original Donnachie--Landshoff fit (dotted), the geometrical
model for the proton (solid),
and the impact factors of \protect\cite{Fukugita:1978fe} 
(FK, long--dashed) and \protect\cite{Levin:gg} 
(LR, short--dashed), figure from \protect\cite{Dosch:2002ai} 
\label{fig:allcurves}}
\end{figure}
Figure \ref{fig:allcurves} shows that the parameters of all models 
for the Odderon--proton coupling can be chosen in such a way that 
the description of the data is as good as with the original DL fit 
(and a better description can hardly be expected within that 
framework). 
One finds that the optimal values for the coupling constant in 
the impact factors should be chosen relatively small, 
in the range $\alpha_s \simeq 0.3 - 0.5$. In the geometric model 
of the proton one finds that a good description of the data 
requires a relatively small diquark cluster in the proton, 
with a size of less than $0.35\,\mbox{fm}$. 

It is worth noting here that 
originally larger values had been proposed for $\alpha_s$ 
in the impact factors in 
\cite{Fukugita:1978fe} and \cite{Levin:gg}. In the case of 
\cite{Fukugita:1978fe} even $\alpha_s=1$ was suggested, 
and that value has also been used in a number of predictions 
for other processes involving momentum scales similar to 
the value of $\sqrt{-t}$ in the dip region. Since the cross section 
for these processes depends on $\alpha_s$ very strongly, 
like $\alpha_s^6$, the original choice for its value is likely to 
overestimate the cross section considerably. 
Independent of those two models for the Odderon--proton 
coupling we make a seemingly trivial but very important 
observation here. The cross section for processes with 
an Odderon exchange depend very strongly on $\alpha_s$, 
namely with a high power. Therefore even a small 
uncertainty in the choice of the correct value (or momentum 
scale) for $\alpha_s$ leads to the largest uncertainty of 
the prediction of the cross section. This significantly contributes 
to the difficulties in finding the Odderon. 

Also the so--called $\rho$-parameter 
\be
\rho (s) = \frac{\real A(s,t=0)}{\imag A(s,t=0)}
\ee
was for some time considered to be an observable which would 
be suitable for finding the Odderon. A sign of the Odderon 
could be a nonvanishing difference of the $\rho$-parameters 
for $pp$ and $p\bar{p}$ scattering, 
\be
\Delta \rho (s)= \rho\,^{p\bar{p}}(s) - \rho\,^{pp}(s)
\ee
at high energies. Unfortunately a precise determination of the 
$\rho$-parameter is extremely difficult experimentally. 
According to the latest measurement by the UA4/2 collaboration 
at the CERN SPS the value of $\Delta \rho$ is compatible 
with zero \cite{Augier:1993sz}. 
However, it should be emphasized that even $\Delta \rho = 0$ would 
not exclude an Odderon, but would only rule out specific models for 
the soft Odderon. In total, the situation of the $\rho$-parameter 
is not really conclusive. 

In the processes we have discussed so far the Odderon 
gives only one among many contributions to the cross section 
and is hence rather difficult to identify, especially because the 
other contributions are of nonperturbative nature and often 
only poorly known. In the past few years there has been 
a change of direction in the search for the Odderon. In order 
to avoid that main problem one now looks for exclusive processes 
in which the Odderon is the only contribution to the 
scattering amplitude, with the possible exception of a reggeon 
that is rather well understood. These processes have in general 
a much smaller cross section. But they have the advantage that 
already their observation would establish the existence of the 
Odderon. 

\begin{figure}
\vspace*{0.4cm}
\begin{center}
\input{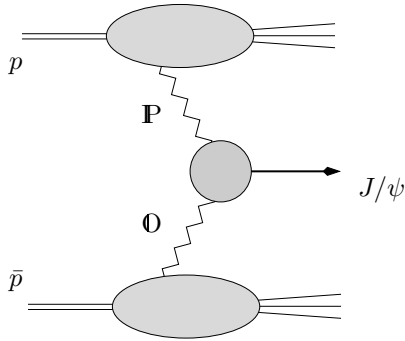}
\end{center}
\caption{Pomeron--Odderon fusion mechanism for double--diffractive
$J/\psi$ production in $p\pbar$ scattering
\label{pomoddfusion}}
\end{figure}

\begin{figure}[ht]
\begin{center}
\input{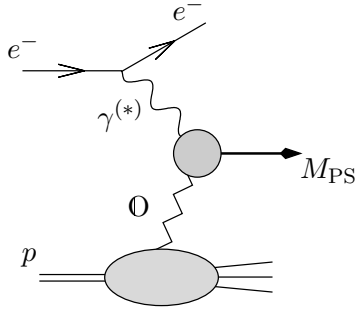}
\end{center}
\caption{Diffractive production of a pseudoscalar meson in $ep$ scattering
\label{diffmesfig}}
\end{figure}

A process of this kind are for example the double 
diffractive production of vector mesons in $pp$ collisions, 
$p\,p \to p\,p \,M_V$, see figure \ref{pomoddfusion}, 
in which the vector meson would be produced via a Pomeron--Odderon 
fusion mechanism. 
This process has been considered for example in \cite{Schafer:na}. 
Another interesting possibility is the diffractive production of 
pseudoscalar or tensor mesons in $ep$ collisions due to 
$\gamma^{(*)}p \to p\, M_{PS/T}$, as illustrated in 
figure \ref{diffmesfig}. In both processes 
the quantum numbers of the produced mesons require 
an Odderon exchange. 

\begin{figure}
\vspace*{0.4cm}
\centering
\includegraphics[width=0.75\textwidth,clip]{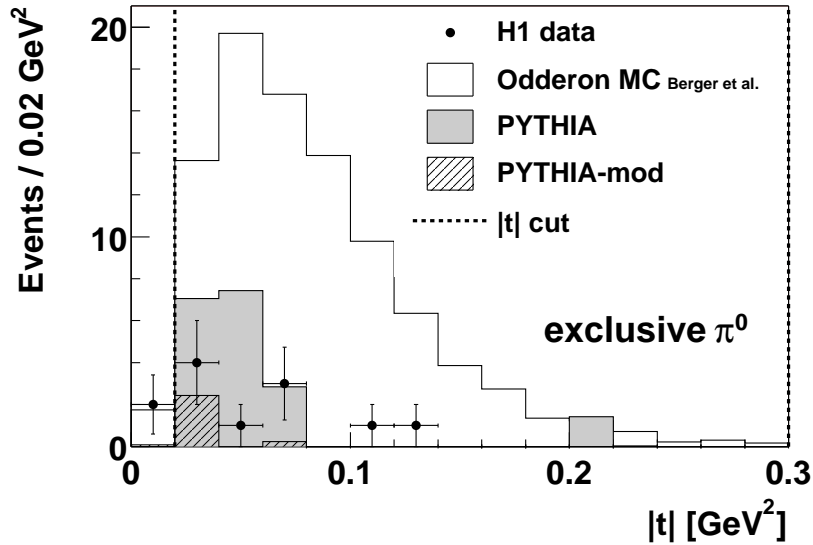}
\caption{Measured $t$-distribution for Odderon candidate 
events with $M_{\gamma \gamma}< 335\,\mbox{MeV}$; 
figure from \protect\cite{Adloff:2002dw}
\label{h1fig}}
\end{figure}

Let us consider in some more detail the diffractive production of 
pseudoscalar mesons. The largest cross sections are clearly to be 
expected for the photoproduction of light mesons. Here both the 
real photon as well as the light mesons require a completely 
nonperturbative treatment. This process has therefore been 
studied in the framework of Regge theory in \cite{Kilian:1997ew}, 
and using more sophisticated methods in \cite{Berger:1999ca}. 
The latter approach makes use of an implementation 
\cite{Kramer:tr,Kraemer:rc,Dosch:1992cu,Dosch:1994ym}
of the stochastic vacuum model (SVM) of 
\cite{Dosch:1987sk,Dosch:ha,Simonov:1987rn} in the 
framework of a nonperturbative framework for the description 
of high energy scattering \cite{Nachtmann:1991ua}. That 
approach has proven to be very successful in the case of 
scattering processes mediated by Pomeron exchange. 
One should again expect a suppression of the 
Odderon--proton coupling due to a potential diquark clustering within 
the proton. This can be avoided if one considers only the case in 
which the proton breaks up. 
The resulting estimate for pion production in the process
$\gamma p \to \pi^0 N^*$ at HERA is $\sigma \simeq 200\,\mbox{nb}$. 
A similar estimate for the production of $f_2$ tensor mesons in 
$\gamma p \to f_2 X$ at HERA is $\sigma \simeq 21\,\mbox{nb}$ 
\cite{Berger:2000wt}. 
Given these numbers there seemed to be a realistic chance 
of observing the Odderon in these processes at HERA. 
Figure \ref{h1fig} shows the rather disappointing result of a 
measurement of diffractive pion production \cite{Adloff:2002dw}. 
The data points are far below the expectations of a Monte Carlo 
based on the results of \cite{Berger:1999ca}, and are well 
compatible with the expected background. The 
calculations of \cite{Berger:1999ca} rely on nonperturbative 
techniques and naturally have a large uncertainty. Nevertheless, 
the rather dramatic failure of the prediction in this case is quite 
puzzling and is not yet understood. Possible reasons might be 
an extremely low intercept of the soft Odderon, 
a suppression of the Odderon--proton coupling due 
to some reason possibly involving some assumptions made in 
the MSV, or finally a suppression of the Odderon--pion coupling 
possibly related to the special role of the pion as a Goldstone boson. 
The latter possibility can in principle be tested by measuring the 
corresponding cross sections for other pseudoscalar or tensor mesons. 
Most probably we are missing an important insight here, and 
further study is urgently needed. 

Less uncertainties are involved if one considers the diffractive 
production of heavy pseudoscalar mesons like the $\eta_c$. 
The mass of the charm quark provides a large scale and hence 
one can approach this process using perturbation theory 
\cite{Czyzewski:1996bv,Engel:1997cg,Bartels:2001hw}. 
In the first two references a simple three--gluon exchange 
is used for the Odderon without resummation, whereas in 
\cite{Bartels:2001hw} the resummed BLV Odderon solution 
is used. The expected cross sections for the photoproduction 
of $\eta_c$ range up to about $50\,\mbox{pb}$, and even 
these values are obtained only with a very optimistic choice 
of $\alpha_s$ in the impact factors coupling the Odderon 
to the proton, see the discussion above. More realistically one 
should expect a factor 30 less. But even the optimistic estimate 
would be much to small for a realistic chance of observing 
this process at HERA. 

The cross sections for the diffractive processes discusses above 
contain the square of the Odderon amplitude and thus have 
an enhanced sensitivity to the uncertainties of this amplitude 
like the coupling of the Odderon to external particles etc. 
Interestingly it is also possible to find observables which are 
only linear in the Odderon amplitude. If one considers final 
states which can be produced both via Pomeron and Odderon 
exchange there can be Pomeron--Odderon interference effects 
\cite{Brodsky:1999mz} which typically occur in suitably 
constructed asymmetries. On the parton level these asymmetries 
usually vanish, but in the process of hadronization there occur 
additional Breit--Wigner phases which can give rise to a sizable 
effect in such asymmetries. The violation of quark--hadron 
duality is therefore crucial for these observables. 
Probably the most favorable observable of this kind is 
the charge asymmetry in diffractive $\pi^+ \pi^-$ production 
in $ep$ collisions at HERA. The charge asymmetry is defined as 
\be
{\cal A} (Q^2,t,m_{\pi^+\pi^-}^2) =
\frac{ \int \cos \theta \,d\sigma(s,Q^2,t,m_{\pi^+\pi^-}^2,\theta)}{ \int 
d\sigma(s,Q^2,t,m_{\pi^+\pi^-}^2,\theta)}
\,,
\ee
where $\theta$ is the polar angle of the $\pi^+$ in the dipion rest frame. 
A pion pair can be produced both in a $C$-odd and in a $C$-even state, 
the former via Pomeron exchange and the latter via 
Odderon exchange. The charge asymmetry is constructed such that it is 
given by the interference term of the Pomeron and the Odderon amplitude. 
One expects the asymmetry to be of the order of about $10 - 20\,\%$ 
for both photoproduction \cite{Ivanov:2001zc,Ginzburg:2002zd} 
and electroproduction \cite{Hagler:2002nh,Hagler:2002nf}. 
A measurement of the charge asymmetry would clearly be a very 
important step in the search for the Odderon. 

\section{Summary}

The perturbative Pomeron and the corresponding rise of 
cross sections are firm predictions of QCD. But as in the 
case of any other approximation in physics one has to 
identify the appropriate range of applicability of the BFKL 
Pomeron. 
Contrary to earlier expectations it turns out that the 
perturbative Pomeron applies only in a limited number 
of experimentally testable scattering processes, and 
even there only in suitable kinematic situations. 
In these, there are in fact some indications for a 
behavior compatible with the perturbative Pomeron. 
The nonobservation of the perturbative Pomeron in 
other processes, on the other hand, should actually  
not cause any excitement. 

If our understanding of high energy scattering based on 
gluon exchanges is correct there should or even must be 
an Odderon of some kind. This holds in particular in 
situations in which perturbation theory is applicable. 
In situations involving lower momentum scales, however, 
it is conceivable that Odderon exchange is suppressed in 
some form, for example due to a very low intercept of the 
nonperturbative Odderon. 

This conference took place in the building of the Heidelberger 
Akademie der Wissenschaften. Besides the warm hospitality 
of the staff of the Akademie we also enjoyed an impressive view 
on the Heidelberg castle. The poetic touch of this place can 
hardly be matched by any scientific statement. Nevertheless, 
I would like conclude with another version of my summary of the 
status of the Odderon that is slightly more poetic than 
the one given above:\footnote{I have chosen the German (that is the local) 
language here since my ability to make rhymes is somewhat  
limited in other languages.}\\

\noindent
Das Odderon, das Odderon\\
Man hat geh\"ort davon. \\
Es beliebt sich sehr zu zieren,\\
will man es detektieren.\\
Nebul\"os sind alle Zeichen,\\
ein Knick in Sigma, der mu{\ss} reichen.\\
Doch da{\ss} die Daten hier sich beugen,\\
vermag so manchen nicht zu \"uberzeugen.\\
Ein Eta, diffraktiv erschienen,\\
w\"urde allerliebst uns dienen.\\
Jedoch, die Suche bleibt sehr schlicht,\\
man findets einfach nicht.\\
Und wenn wir noch so hoffen,\\
am Ende ist wie immer alles offen.

\section*{Acknowledgements}
I would like to thank H.\,G.\ Dosch, O.\ Nachtmann and 
V.\ Schatz for helpful discussions.

\end{document}